\documentclass{epl}
\title{Direct equivalence between quantum phase
transition phenomena in radiation-matter and magnetic systems:
scaling of entanglement}
\shorttitle{Direct equivalence between quantum phase
transition etc.}
\author{J. Reslen\inst{1}\thanks{E-mail:\email{j-reslen@uniandes.edu.co}} \and L. Quiroga\thanks{E-mail:\email{lquiroga@uniandes.edu.co}} \inst{1} \and N. F. Johnson\inst{2}\thanks{E-mail:\email{n.johnson@physics.ox.ac.uk}}}
\institute{
  \inst{1} Departamento de F\'{\i}sica, Universidad de Los Andes, A.A.4976,
Bogot\'a D.C., Colombia \\
  \inst{2} Centre for Quantum Computation and Department of Physics,
University of Oxford, Clarendon Laboratory, Parks Road, OX1 3PU, U.K
}
\shortauthor{J. Reslen \etal}

\begin{document}

\maketitle

\begin{abstract}
We show that the quantum phase transition arising in a
standard radiation-matter model (Dicke model) belongs to the same
universality class as the infinitely-coordinated, transverse field
XY model. The effective qubit-qubit exchange
interaction is shown to be proportional to the square of the
qubit-radiation coupling. A universal finite-size scaling is derived for the
corresponding two-qubit entanglement (concurrence) and a
size-consistent effective Hamiltonian is proposed for the qubit subsystem.
\end{abstract}

Quantum phase transitions (QPTs) are associated with a dramatic
change in the physical properties of a system at zero temperature
when a parameter varies around its critical value. It is
well-known that very different systems can exhibit similar
behavior in this critical regime, giving rise to the concept of
{\it universality}. Enlarging a given universality class by the
addition of systems from very different areas of physics, is a
very important step toward unifying our understanding of the basic
physics underlying apparently disconnected complex phenomena.
Recently there have been studies of light-controlled condensed
matter systems displaying QPTs with atoms in extreme
one-dimensional confinements \cite{belen}, ions driven by properly
tuned and pulsed light \cite{porras} and fermionic atoms in
optical superlattices \cite{belen2}. Fully quantum mechanical
models of radiation-matter systems are also being considered, and
are important for several reasons: Scalable and distributed
quantum information processing (QIP) devices will demand the
integration of matter quantum bits (qubits) such as atoms, trapped
ions, semiconductor quantum dots or SQUIDs with photons. In
addition, the capability of photons to control and modify the
coupling between physically distant qubits makes them appropriate
for manipulating and transferring quantum information.

QPTs themselves are beginning to be studied using concepts from
quantum information theory. Special attention has been paid to the
connections between subsystem entanglement and critical phenomena
\cite{fazio,nielsen,jvidal}. There is an obvious correspondence
between condensed matter magnetic systems comprising $N$ spin-1/2
particles, and quantum registers with $N$ qubits. Most of the
quantum information-based analysis on QPTs performed to date
corresponds to critical magnetic systems where the exchange
coupling between spins is fixed from the outset by Nature. However
there is also interest in studying QPTs where the spin/qubit
couplings can be controlled externally, using light for example.
One of the most famous radiation-matter systems giving rise to a
QPT is the Dicke model \cite{dicke,hioe}. In this model, $N$
two-level systems (qubits) are coupled to a single radiation mode
(cavity). Although no direct coupling exists between qubits, a
superradiant phase arises as the qubit-cavity coupling strength
increases. Quantum chaos and entanglement in the Dicke model have
been studied recently using spin-boson transformation techniques
\cite{brandes1,brandes2}. However a full understanding of the
entanglement (concurrence) finite-size scaling behavior and the
associated universality class, are still lacking.

Here we show that the reduced dynamics of a qubit system embedded
in a photon field, as described by the Dicke model, displays
magnetic-like behavior and corresponds to the same universality
class of quantum critical phenomena as an infinitely-coordinated
ferromagnetic system. The qubit-cavity detuning determines an
effective magnetic field while the square of the qubit-cavity
coupling strength yields the effective qubit-qubit `exchange'
interaction. As a result, we are able to establish an unambiguous
quantitative connection between the associated scaling exponents
and effective qubit Hamiltonian in two quite different physical
systems. Although Lambert et al. \cite{brandes2} have already
noted similarities between the Dicke and collective XY model in
terms of scaling exponents for the concurrence, the present work
establishes a systematic link between these models and lays these
similarities on a firm theoretical footing.

We are interested in a system formed by $N$ identical two-level atoms (qubits)
coupled to a single-mode cavity, described by the Hamiltonian
\begin{equation}
\label{Eq:sys}
H=a^{\dag}a+\frac {\epsilon}{2} \sum_{i=1}^N \sigma_{i,z}+
\frac {\lambda}{\sqrt{N}} \sum_{i=1}^N (a^{\dag}+a)(\sigma_i^{\dag}+\sigma_i)
\end{equation}
where $a^{\dag}$ ($a$) is the creation (annihilation) operator
for single-mode cavity photons, $\sigma_{i,z}=|1_i\rangle \langle
1_i|-|0_i\rangle \langle 0_i|$, $\sigma_i^{\dag}=|1_i\rangle
\langle 0_i|$, $\sigma_i=|0_i\rangle \langle 1_i|$ with
$|0_i\rangle$ and $|1_i\rangle$ $(i=1,2,...,N)$ being the ground
and excited states of the $i$'th qubit, respectively, and $\lambda$ is the
qubit-cavity coupling. Note that we
do not make a rotating wave approximation. We take the energy of a
confined photon as the unit of energy. Resonance between qubit
energy levels and photons corresponds to $\epsilon=1$. This
Hamiltonian can also be written as
\begin{equation}
\label{Eq:jsys}
H=a^{\dag}a+\epsilon J_z+\frac {2\lambda}{\sqrt{N}} (a^{\dag}+a)J_x
\end{equation}
where $J_z=\frac {1}{2} \sum_{i=1}^N \sigma_{i,z}$ and
$J_x=\frac {1}{2}\sum_{i=1}^N (\sigma_i^{\dag}+\sigma_i)$. We
denote this latter Hamiltonian as $H=H_0+V$ where
$H_0=a^{\dag}a$ and $V=\epsilon J_z+ \frac {2\lambda}{\sqrt{N}}
(a^{\dag}+a)J_x$.

We shall be concerned with physical qubit quantities such as `magnetization'
and entanglement. These quantities can be represented by a generic
qubit operator $Q$. Hence in thermodynamic equilibrium at inverse temperature
$\beta=(k_B T)^{-1}$ any qubit expectation value can be expressed as
\begin{eqnarray}
\label{Eq:q}
<Q>=\frac {Tr_{qb}[QTr_{ph}[e^{-\beta H}]]}{Tr_{qb}
[Tr_{ph}[e^{-\beta H}]]}=
\frac {Tr_{qb}[Qe^{-\beta H_{qb}}]}{Tr_{qb}[e^{-\beta H_{qb}}]}
\end{eqnarray}
where $Tr_{qb}$ ($Tr_{ph}$) denotes a trace over the qubit
(cavity) subsystem and the last equality is the formal definition
of an effective qubit Hamiltonian $H_{qb}$. In order to eliminate
the cavity degrees of freedom and obtain a size-consistent
$H_{qb}$ (i.e. a reduced qubit Hamiltonian which scales with the
system's size) we follow the procedure sketched out by Polatsek
and Becker \cite{becker}. The resulting effective qubit
Hamiltonian can be written as
\begin{equation}
\label{Eq:eff1}
H_{qb}=-\frac {1}{\beta} {\rm ln}\langle e^{-\beta(L_0+V)}
\rangle_{ph}
\end{equation}
where the thermal averaging is carried out with respect to the cavity degrees of freedom, i.e.
$\langle O\rangle_{ph}=Tr_{ph}(Oe^{-\beta a^{\dag}a})/Tr_{ph}(e^{-\beta
a^{\dag}a})$;
$L_0$ is the Liouville operator associated with $H_0$ and defined by
$L_0O=[a^{\dag}a,O]$ for any operator
$O$. A cumulant expansion is well-suited to calculating a
size-consistent $H_{qb}$, yielding
\begin{equation}
\label{Eq:cum1}
H_{qb}=-\frac {1}{\beta} \sum_{n=1}^\infty (-1)^n \frac
{\beta^n}{n!}H_n
\end{equation}
where $H_n$ represents the qubit-operator corresponding to the
$n$'th cumulant. Cumulants coming from eq.~(\ref{Eq:eff1}) are
themselves operators which act on qubit states, hence any
expression involving cumulants must be properly symmetrized. The
lowest-order cumulants are:
\begin{equation}
\label{Eq:m1}
H_1=\epsilon J_z \hspace {0.3 cm} , \hspace {0.3 cm} H_2=\bigg[\frac {2\lambda}{\sqrt{N}}\bigg]^2
(2h(\beta)+1)J_x^2
\end{equation}
%
%
where $h(\beta)=(e^{\beta}-1)^{-1}$ is the Bose factor which
determines the average photon number in an isolated cavity at
temperature
$T$. Due to the superradiance phenomenon which occurs in the Dicke
model, the actual photon number has very different behavior
on either side of the quantum and/or thermal phase
transition. It can be seen that
$H_1$ corresponds to the non-interacting qubit system while $H_2$ gives
rise to
the lowest order effective qubit-qubit coupling.

Every cumulant contains a term of the form
\begin{equation}
\label{Eq:sm1}
\langle VL_0^nV\rangle_{ph}=\bigg[\frac
{2\lambda}{\sqrt{N}}\bigg]^2 \big[(1+(-1)^n)h(\beta)+1\big]J_x^2
\end{equation}
which enables us to perform a closed-form summation in
eq.~(\ref{Eq:cum1}) over an infinite set of contributing terms. Up
to second order in $V$, this renormalization summation yields a
temperature-dependent effective Hamiltonian given by
\begin{equation}
\label{Eq:h2}
H_{qb}^{(2)}(\beta)=\epsilon  J_z-\bigg[\frac
{2\lambda}{\sqrt{N}}\bigg]^2 \big[1+\frac{2}{\beta(h(\beta)+1)}\big]J_x^2
\end{equation}
At zero temperature (i.e. $\beta\longrightarrow\infty$) this
effective Hamiltonian is a special case of that studied in
ref.~\cite{jvidal} corresponding to a ferromagnetic
infinitely-coupled XY model in a transverse magnetic field with a
QPT at $2\lambda_c=1$ in the thermodynamic limit (i.e.
$N\longrightarrow\infty$). Our analysis therefore provides a novel
demonstration of the well-known QPT arising in the Dicke model at
$\lambda_c=1/2$ \cite{hioe}. The broken symmetry associated with
the phase transition is related to parity for the Dicke model as
given by $e^{i\pi (a^{\dagger}a + J_z + \frac{N}{2})}$, while for
the effective $H_{qb}^{(2)}$ model this becomes $e^{i\pi (J_z +
\frac{N}{2})}$. In contrast to a previously employed continuous
variable representation \cite{brandes1,brandes2}, this effective
anisotropic XY model Hamiltonian $H_{qb}^{(2)}$ provides an ideal
starting point for studying the QPT of the Dicke model since we
can follow the qubit subsystem's quantum-mechanical evolution as a
function of the effective qubit-qubit coupling. Furthermore, this
mapping onto the infinitely-coordinated XY model which results
from our approach, yields a deep understanding of the scaling
properties for finite $N$ as well as the mean-field behavior of
the Dicke model in the thermodynamic limit.
\begin{figure}[h]
\twofigures[width=0.35\textwidth, angle=-90 ]{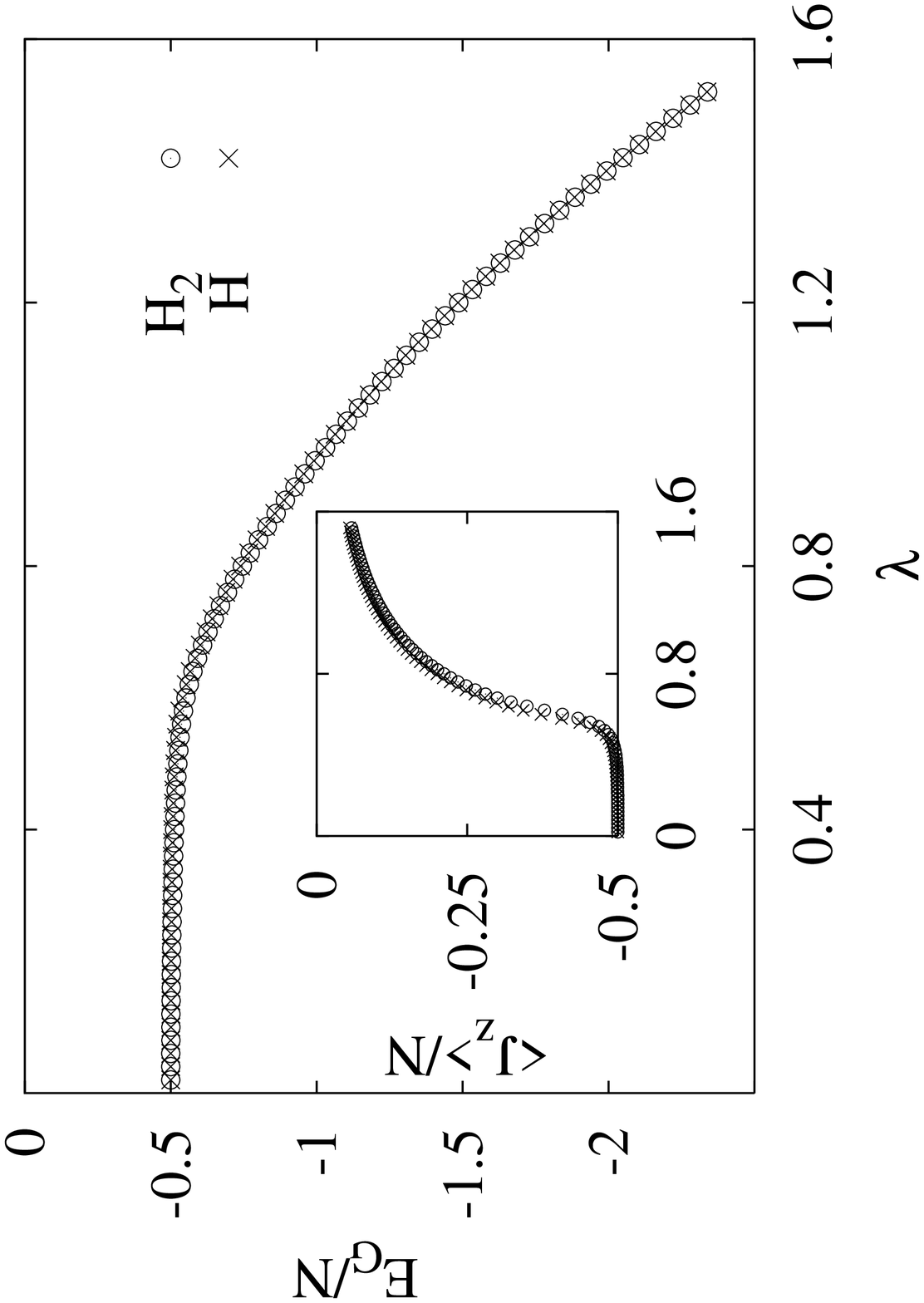}{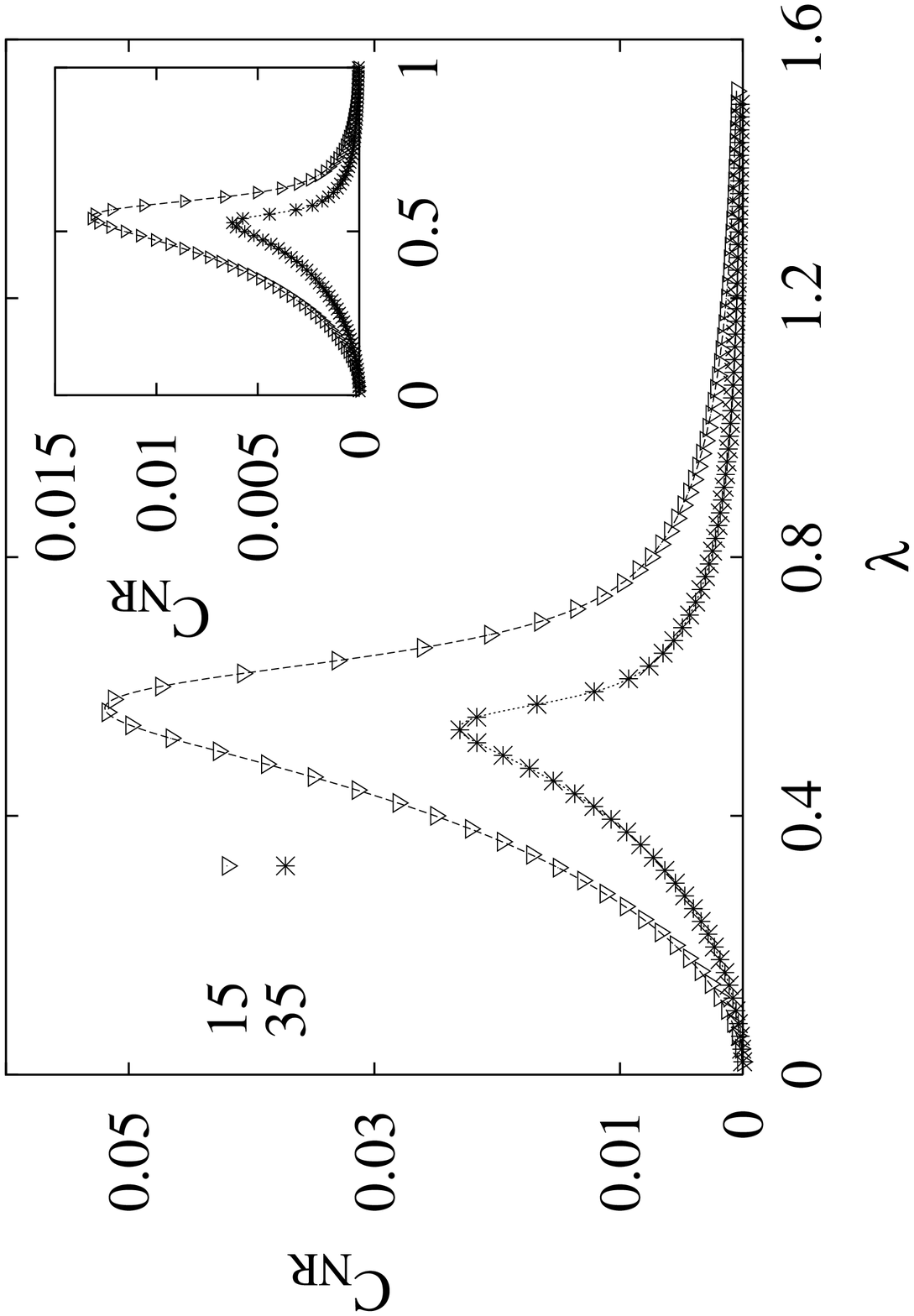}
\caption{Ground state energy and z-magnetization (inset) as a
function of $\lambda$ for $N=15$ qubits, as obtained from the full
Dicke Hamiltonian ($\times$) and from the effective Hamiltonian
($\circ$).}
\label{fig:1a}
\caption{Non-rescaled concurrence as a function of $\lambda$
for the XY model and for the full Dicke Hamiltonian (inset) for
$N$=15 and 35.}
\label{fig:1b}
\end{figure}
\begin{table}
\caption{Critical exponent $b$ at zero temperature for the
finite-$N$ scaling of the x-magnetization at $\lambda_c$ (first
row), z-magnetization at $\lambda_c$ (second row), difference
between the concurrences at $\lambda_c$ for the thermodynamic
limit and for $N$ qubits (third row), difference between the
maxima of the concurrences for the thermodynamic limit and for $N$
qubits (fourth row), and the difference between the coupling
strength for $N$ qubits at maximum concurrence and $\lambda_c$
(fifth row). $N\leq 35$ for the Dicke model and $100\leq N\leq
500$ for the anisotropic XY model.}
 \label{tab:1}
\begin{center}
\[
\begin{array}{|c|c|c|}  \hline f(N)\sim N^b & Dicke & XY  \\ \hline
\frac{\sqrt{\langle J_x^2
\rangle_{\lambda_c}}}{N} & -0.35 \pm 0.01 & -0.34 \pm 0.01  \\
\hline \frac{1}{2}-\left| \frac{\langle J_z
\rangle_{\lambda_c}}{N} \right| & -0.54 \pm 0.01& -0.55 \pm 0.01\\
\hline C_\infty(\lambda_c)- C_N(\lambda_c) & -0.26 \pm 0.01& -0.30
\pm 0.01\\ \hline C_\infty^M-C_N^M & -0.28 \pm 0.03& -0.30 \pm 0.03\\
\hline
 \lambda_N^M - \lambda_c & -0.65 \pm 0.03& -0.66 \pm 0.03\\ \hline
\end{array}
\]
\end{center}
\end{table}
\begin{figure}[h]
\twofigures[width=0.35\textwidth, angle=-90]{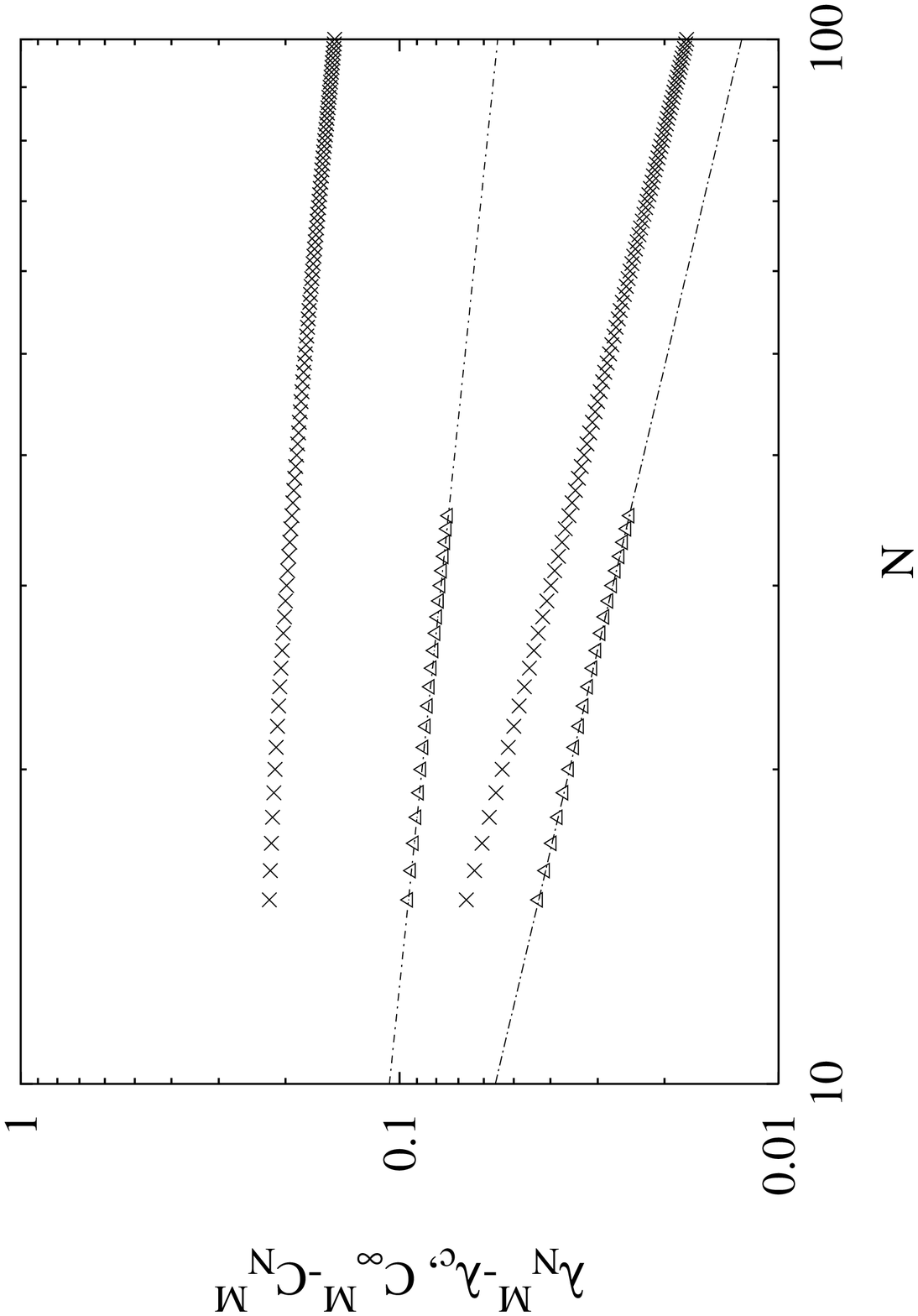}{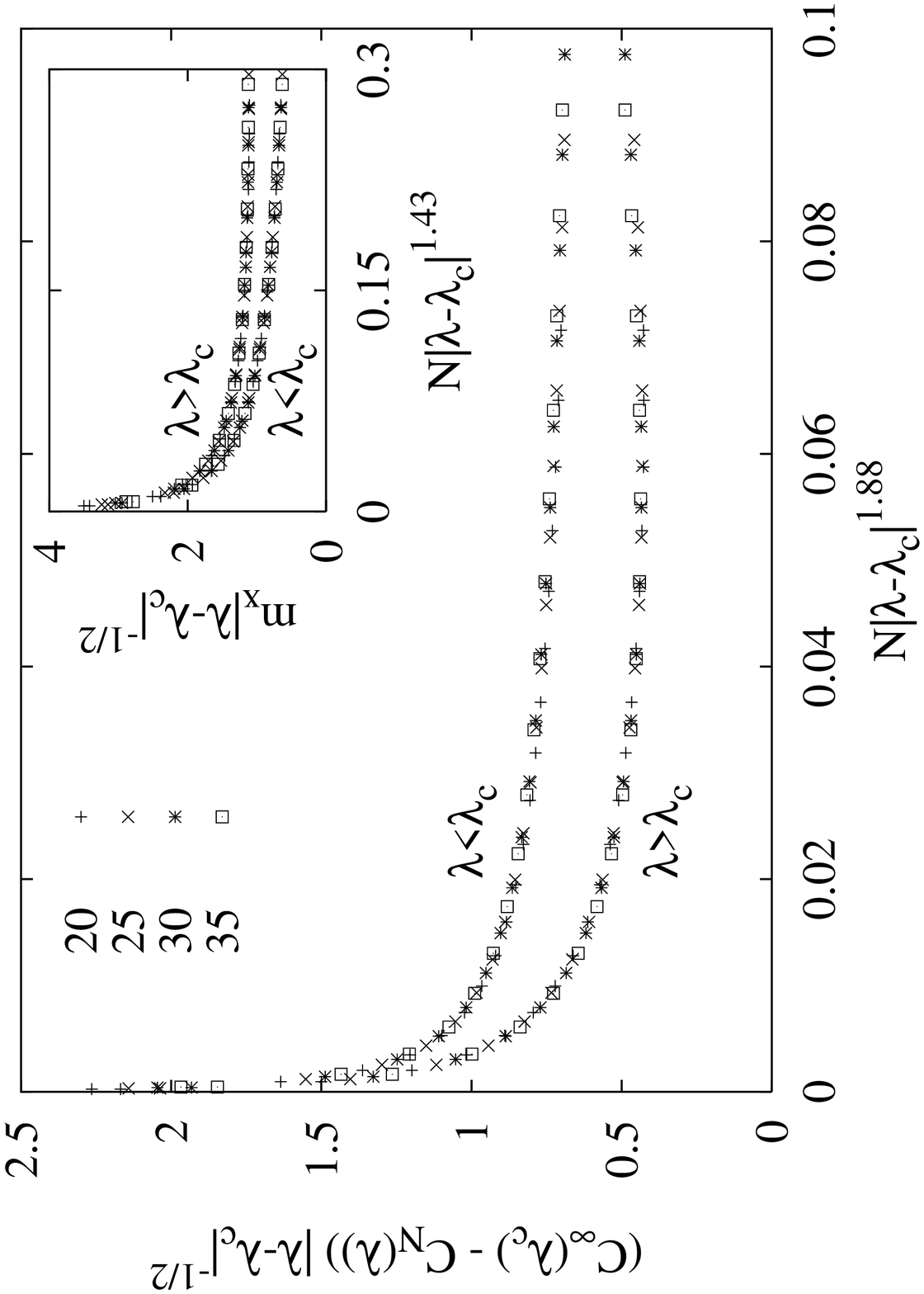}
\caption{Dependence of $\lambda_N^M - \lambda_c$ (lower curves)
and $C_\infty^M-C_N^M$ (upper curves) on $N$ for $H$ ($\triangle$)
and $H_{qb}^{(2)}$ ($\times$). $\lambda_N^M$ is the value of
$\lambda$ for which the concurrence is maximum, while $C_N^M$ is
the maximum of the rescaled concurrence.}
\label{fig:2}
\caption{Finite-size scaling for the rescaled concurrence in the
Dicke model. Points corresponding to different $N$ collapse on the
same curve. Inset: Finite-size scaling for the x-magnetization
$m_x=\frac{\sqrt{\langle J_x^2 \rangle}}{N}$. }
\label{fig:3}
\end{figure}
Fig.~\ref{fig:1a} shows the ground-state energy (at zero temperature) of
a Dicke system with $N$ qubits, as a function of the
qubit-radiation coupling $\lambda$. For simplicity, we only
consider  the case of resonance from now on, i.e. $\epsilon=1$.
For comparison, the ground-state energy produced by the effective
Hamiltonian in eq.~(\ref{Eq:h2}) is also included in the plot. The
agreement between both ground-state energies is excellent even for
low $N$. The qubit subsystem's `z-magnetization', which is the order
parameter for this QPT, undergoes a quantum phase transition at
$\lambda_c=1/2$ as seen in fig.~\ref{fig:1a} (see inset) for both
the exact Dicke model and the effective Hamiltonian. This indicates that the
physics of the QPT in the Dicke model is indeed captured by the effective
Hamiltonian of eq.~(\ref{Eq:h2}).

As a more stringent test of the effective Hamiltonian in
eq.~(\ref{Eq:h2}),  we now compare other thermodynamic as well as quantum
features. An important ground-state quantum feature is the entanglement between
two qubits. We adopt the concurrence
$C_{NR}$ as a measure of this entanglement
\cite{wooters}. We also study the {\it rescaled concurrence}
$C=NC_{NR}$ in order to obtain non-trivial information in the thermodynamic
limit where $C_{NR}=0$. In contrast with thermodynamic results that only
involve the calculation of the partition function and averages obtained from
it, the concurrence probes the internal structure of the
ground-state wavefunction in a more detailed way. In fig.~\ref{fig:1b} we display the
evolution of the concurrence as a function of $\lambda$, for both the exact
Dicke model and the effective Hamiltonian of eq.~(\ref{Eq:h2}).
The qualitative form of the concurrence for both models
is very similar, even though the effective Hamiltonian overestimates the
maximum value. This overestimation is easily understood: the
effective Hamiltonian drops higher-order interaction terms which incorporate
additional interactions among the qubits and would hence
modify the ground-state
wavefunction. In the thermodynamic
limit, the rescaled concurrence displays a singularity at
$\lambda=\lambda_c$ in agreement with the findings in refs.~\cite{jvidal,brandes2}. However, at
$\lambda_c$ the rescaled concurrence goes to 1 for the XY model while it goes
to $1-\frac{\sqrt{2}}{2}$ for the Dicke model.

As a final test of the reduced Hamiltonian of eq.~(\ref{Eq:h2}), we
consider the
finite-size scaling of the `magnetization' as well as the concurrence. Results
can be compared with those reported in ref.~\cite{botet} and show
excellent agreement between the predictions of the Dicke model and the
effective Hamiltonian.

At zero temperature, the energy and the z-magnetization are
identical for both systems in the thermodynamic limit. Physical
quantities depending on the ground state structure, such as the
x-magnetization and concurrence, display some differences.
However, the scaling exponents agree for both models as can be
seen in fig.~\ref{fig:2} and table~\ref{tab:1}. Even for small $N$
in the Dicke model, the quantities $\lambda_N^M - \lambda_c$ and
$C_\infty^M-C_N^M$ can be seen to lie on well-defined straight
lines. For the XY model, large $N$ results lie on straight lines
which are parallel to the Dicke model results. The Dicke model
results approach the thermodynamic limit much faster than those
for the XY model. This is directly related to the fact that terms
have been thrown away in eq.~(\ref{Eq:h2}) which include
interactions between three and more spins, thereby enhancing the
mean-field nature of this system.

The excellent agreement between the critical exponents for the two
models provides clear evidence that they both belong to the same
universality class. It is well known that in both Hamiltonians,
each qubit is coupled to every other qubit with the same strength
and hence no typical lengthscale can be associated with them.
Botet et al.\cite{botet} proposed relating the
infinitely-coordinated magnetic system to short-range interaction
models, and hence were able to perform a systematic study of
finite-size scaling functions for the x-magnetization and gap
energy in the case of a Hamiltonian such as $H_{qb}^{(2)}$.
Following this line of reasoning, we show in fig.~\ref{fig:3}
that this scaling hypothesis is valid for the Dicke model as well.
In particular there exist scaling functions in the critical region
$|\lambda-\lambda_c|\rightarrow 0$: the concurrence behaves as
$C_{\infty}(\lambda_c)-C_N(\lambda)=|\lambda-\lambda_c|^{a_{mf}}F_c(N|
\lambda-\lambda_c|^\nu)$ with $F_c$ a function of $N|
\lambda-\lambda_c|^\nu$ only. When $x\rightarrow \infty$,
$F_c(x)\rightarrow constant$ and
$C_{\infty}(\lambda_c)-C_{\infty}(\lambda)\sim |\lambda -
\lambda_c|^{a_{mf}}$ while for $x\rightarrow 0$,
$F_c(x)\rightarrow x^{-a_{mf}/\nu}$ and
$C_{\infty}(\lambda_c)-C_N(\lambda_c)\sim N^{-a_{mf}/\nu}$. In
ref.~\cite{reslen} the concurrence in the XY model
($N\longrightarrow \infty$, eq.~(\ref{Eq:h2})) has been explicitly
calculated yielding
\begin{equation}
C(\lambda)=
\begin{array}{c}
1-\sqrt{1-\left(\frac{\lambda}{\lambda_c}\right)^2} \hspace {0.3 cm} , \hspace {0.3 cm} 0 < \lambda < \lambda_c \\
1-\sqrt{1-\left( \frac{\lambda_c^2}{\lambda^2} \right)^2} \hspace {0.3 cm} , \hspace {0.3 cm}
\lambda_c < \lambda < \infty \\
\end{array}
\end{equation}
with $\lambda_c=\frac{\sqrt{\epsilon}}{2}$. This implies that,
close to $\lambda_c$,
$1-C\rightarrow|\lambda-\lambda_c|^{\frac{1}{2}}$. If we assume
for the moment that both models, Dicke and infinitely coordinated
XY, belong to the same universality class, we can then borrow the
result $a_{mf}=\frac{1}{2}$ from the XY concurrence and apply it
to the Dicke concurrence. This exponent has also been found for
the full Dicke model in ref.~\cite{brandes2} using a spin-boson
transformation scheme. The exponent $\frac{a_{mf}}{\nu}$ has been
numerically obtained (see table~\ref{tab:1}) yielding $\nu=1.88$.
The collapse of all the data on a single curve for each region
(coupling below or above the critical coupling strength) is a
verification of the existence of scaling behavior of the
concurrence in the Dicke model. In fig.~\ref{fig:3} (inset) the
x-magnetization multiplied by $|\lambda - \lambda_c|^{-1/2}$, is
plotted as a function of $N|\lambda - \lambda_c|^{1.43}$. The
critical exponent in the argument of the scaling function is
$1.43$, which is only slightly smaller than that found for the
same exponent in the XY model case ($1.5$). This similarity in
scaling exponents reinforces the main message of this paper, that
the scaling behaviour of the Dicke model is intimately related to
that of the anisotropic infinitely-coordinated XY model.

In summary, we have shown that a radiation-matter (i.e. Dicke)
model can be mapped on to an infinitely-coordinated spin/qubit
system with an effective magnetic field determined by the detuning
between qubit and cavity, and a long-range exchange interaction
given by the square of the qubit-cavity coupling strength. An
immediate and interesting consequence is that the exchange
interaction can be controlled experimentally simply by changing
the system's temperature or equivalently the mean number of
photons in the cavity. We have also shown that the Dicke model
pertains to the same universality class as other
infinitely-coordinated systems, such as the anisotropic XY model
in a transverse field. We have demonstrated that the finite-size
scaling hypothesis works well for the Dicke model, and we
therefore postulate that it is also likely to work well for a
range of generalized Dicke models corresponding to a diverse range
of radiation-matter experimental setups. Temperature-dependent
phenomena can also be studied using our size-consistent, effective
qubit-qubit system approach.

\acknowledgments

We thank A.Olaya-Castro for her valuable comments. This work was supported by
the Faculty of Sciences (U. de los Andes), COLCIENCIAS (Colombia) project
1204-05-13614 and DTI-LINK (UK).


\end{document}